\documentclass[letterpaper, 10 pt, conference]{ieeeconf}  

\IEEEoverridecommandlockouts                              
\usepackage[utf8]{inputenc}
\usepackage[T1]{fontenc}
\usepackage{acronym}
\usepackage{graphicx}
\usepackage{multirow}
\usepackage{microtype}
\usepackage{makecell}
\usepackage{subcaption,siunitx,booktabs} 
\usepackage{authblk}
\usepackage{subcaption}
\usepackage{caption}
\usepackage{times}
\usepackage{epsfig}
\usepackage{graphicx}
\usepackage{amsmath}
\usepackage{amssymb}
\usepackage{acronym}
\usepackage{microtype}
\usepackage[english]{babel}

\title{\LARGE \bf
Generalisable 3D Fabric Architecture for Streamlined Universal Multi-Dataset Medical Image Segmentation}

\author{Siyu Liu$^{1}$, Wei Dai$^{1}$, Craig Engstrom$^{1}$, Jurgen Fripp$^{2}$, Stuart Crozier$^{1}$, Jason A. Dowling$^{2}$, Shekhar S. Chandra}

\affil[$^{1}$]{School of Information Technology and Electrical Engineering, University of Queensland, Australia}
\affil[$^{2}$]{Australian e-Health Research Centre, CSIRO, Australia}
\begin{document}
\maketitle
\thispagestyle{empty}
\pagestyle{empty}
\begin{abstract}
Data scarcity is common in deep learning models for medical image segmentation. Previous works proposed multi-dataset learning, either simultaneously or via transfer learning to expand training sets. However, medical image datasets have diverse-sized images and features, and developing a model simultaneously for multiple datasets is challenging. This work proposes \ac{FIRENet}, a universal 3D architecture for simultaneous multi-dataset segmentation and transfer learning involving arbitrary numbers of dataset(s). To handle different-sized image and feature, a 3D fabric module is used to encapsulate many multi-scale sub-architectures. An optimal combination of these sub-architectures can be implicitly learnt to best suit the target dataset(s). 
For diverse-scale feature extraction, a 3D extension of atrous spatial pyramid pooling (ASPP3D) is used in each fabric node for a fine-grained coverage of rich-scale image features. 
In the first experiment, \ac{FIRENet} performed 3D universal bone segmentation of multiple musculoskeletal datasets of the human knee, shoulder and hip joints and exhibited excellent simultaneous multi-dataset segmentation performance. When tested for transfer learning, \ac{FIRENet} further exhibited excellent single dataset performance (when pre-training on a prostate dataset),  as well as significantly improved universal bone segmentation performance. The following experiment involves the simultaneous segmentation of the 10 \ac{MSD} challenge datasets. \ac{FIRENet} demonstrated good multi-dataset segmentation results and inter-dataset adaptability of highly diverse image sizes.  
In both experiments, \ac{FIRENet}'s streamlined multi-dataset learning with one unified network that requires no hyper-parameter tuning.

 \end{abstract} 

\acresetall
Deep learning for medical image segmentation is a rapidly evolving field with the potential to enhance disease diagnosis~\cite{diag1} and treatment planning~\cite{tp1}. The effectiveness of deep learning can be largely attributed to its data-driven nature. However, this reliance on data is also a major limitation in medical image segmentation due to data scarcity. Unlike the abundance of large-scale public datasets~\cite{coco, pascal, cityscape} in 2D computer-vision tasks, (expert) labelled medical image datasets are much smaller in quantity due to several factors:

\begin{figure}[h]
     \centering
     \includegraphics[width=0.45\textwidth]{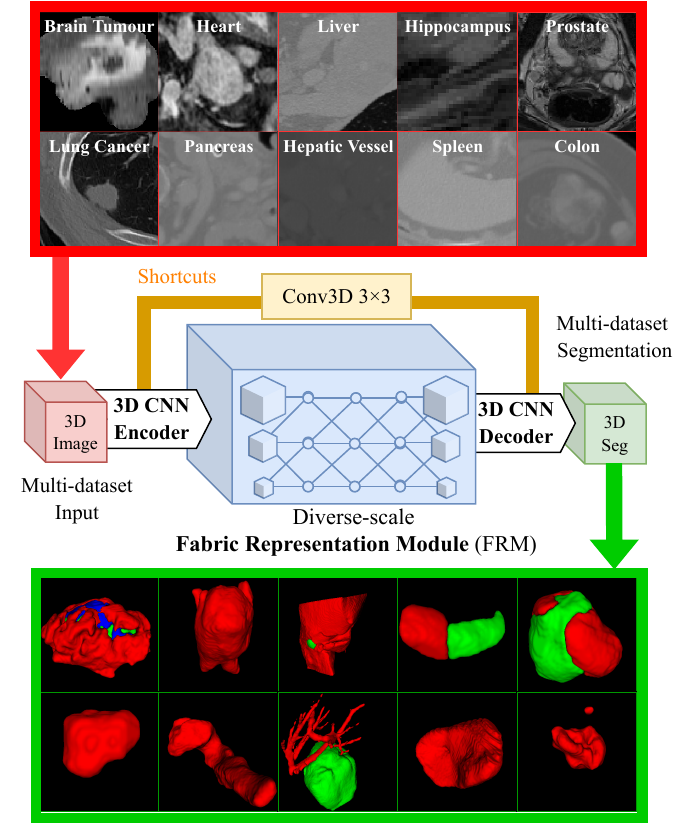}
     \caption{Simultaneous 3D segmentation of the 10 \acf{MSD} challenge datasets using only one instance of \ac{FIRENet}. \ac{FIRENet} uses a the latent fabric module to process diverse-sized features and images. }
     \label{fig:encdec}
\end{figure}

\begin{itemize}
    \item Data acquisition challenges: The acquisition of medical images can be highly specialised and resource-intensive. Voxel-wise manual annotation of 3D volumes (for segmentation) is time-intensive and subject to variable operator error. In addition, careful planning and expert contouring protocols are usually required to minimise intra- and inter-rater segmentation variability.
    
    \item Data fragmentation: Clinical studies involving medical imaging are typically highly focused, relatively small investigations due to high imaging costs. Datasets from different studies can exhibit considerable inter-dataset variations (for example, different imaging fields of view, and dimensions) associated with different acquisition sequences and protocols. Most models lack versatility as they are highly optimised for only one dataset at a time.
    
    \item Access to imaging datasets: Collecting large-scale medical image datasets with expert annotations is difficult as explicit consent, strict adherence to ethics and systematic coordination are required, as the publication of patient data, even when de-identified, is a highly sensitive matter.

\end{itemize}

When facing data scarcity, multi-dataset applications including transfer learning~\cite{transfer, med3d} and simultaneous multi-dataset learning~\cite{3du2net}) have been shown to improve segmentation performance. However, most current deep learning methods for medical image analysis are designed for application on very few datasets. Even architectures that can adapt to different datasets datasets, like nnUNet~\cite{nnunet}, are optimised for one dataset at a time. Works that use one model to aggregate from and transfer features to many datasets remain limited. Currently, existing (and notable) works for simultaneous multi-dataset segmentation in the literature are 3D MDUNet and 3DU$^2$Net, which applied a U-Net-like architecture simultaneously to several \ac{MSD} challenge datasets. However, they only demonstrated limited applicability on a subset of the available \ac{MSD} datasets. To extend the coverage to more datasets with diverse features and sizes, a model with more focus on diverse-scale feature extraction and architecture-level versatility is desirable.

Fabric-based architectures~\cite{grid,interlink,fabric} are an ideal candidate for multi-dataset processing as they create a superposition of many multi-scale sub-architectures.  With this inspiration, this work proposes \ac{FIRENet}, a versatile 3D network that leverages a convolutional fabric for multi-dataset medical image segmentation. The central features of \ac{FIRENet} are:

\begin{enumerate}
    \item For improved adaptation to different medical image datasets, \ac{FIRENet}'s uses a 3D \acf{FRM} to emulate a superposition of many multi-scale sub-architectures. The major advantage of \ac{FRM} is architecture-level generalisability, which is essential as different datasets contain different image and feature sizes.
    
    \item To harness the benefits of multi-scale feature extraction for processing diverse features, \ac{FRM} consists of multi-scale branches of fabric nodes. Furthermore, a 3D extension of \ac{ASPP}~\cite{dlv3} is employed in each \ac{FRM} node to provide fine-grained coverage of different-sized features.
    
    \item The nodes in \ac{FRM} are connected via \ac{TFS} to further optimise \ac{FIRENet}'s multi-dataset adaptability. These connections can be trained to automatically form an architecture from different medical image datasets. (Figure~\ref{fig:encdec}). 
    
    \item \ac{FIRENet}'s generalisable architecture ensures no hyper-parameter modification is required when applying or transferring it to multiple datasets.
    
\end{enumerate}
FIRENet's demonstrated good performance and versatility in various transfer learning and multi-dataset segmentation tasks.
The transfer learning and simultaneous multi-dataset segmentation experiments show FIRENet's versatility in various medical image segmentation tasks, including I) standard single dataset segmentation (during pre-training on an MR prostate dataset) II) simultaneous universal (support for multi-anatomy) bone segmentation with and without transfer learning, and III) simultaneous multi-object segmentation performance for all the 10 image datasets in the \ac{MSD} challenge. Notably, \ac{FIRENet} also exhibits multi-dataset adaptability without tailored training procedures to reach convergence.

\section{Related work}
\subsection{\acp{CNN} for image segmentation}
The most common class of deep learning models for (medical) image segmentation are \acp{CNN}. \acp{CNN} were initially developed for image classification~\cite{cnn, vgg} and were later found suitable for image segmentation tasks. Typically, \acp{CNN} use  consecutive convolution (image filtering) and hierarchical down-sampling to task-related extract features. However, since image segmentation require dense (pixel or voxel-wise) predictions, image segmentation models require feature learning at both global (coarse) and local (fine-detailed) scales. Excessive down-sampling, as used in classic \acp{CNN}~\cite{vgg,resnet} can degrade segmentation accuracy due to loss of image resolution. In response, later works, including UNet~\cite{unet}, incorporated an encoder-decoder architecture that first extracts large-context features using an encoder \ac{CNN}, then re-constructs the full-resolution image using a decoder \ac{CNN} (reversing the down-sampling performed by the encoder). To further aid the decoder in recovering fine details, UNet also makes use of shortcuts for passing high-resolution features directly from the encoder to the decoder. There have been several recent UNet-based networks \cite{unetpp, raunet, reunet, dilatedunet} successfully applied to medical image segmentation tasks.

\subsection{\acp{CNN} for improved architecture generalisability}
Most early-day \acp{CNN} were only suitable for processing limited image and feature sizes, which is not ideal for models intended for complex features or in this case, multiple datasets. In response, multi-scale feature extraction~\cite{mult1,mult2,mult3,mult4,mult5} was introduced and has become a common inclusion in \acp{CNN} for extracting diverse-sized features. Typically, multi-scale feature extraction divides the input into several parallel branches, each with a different receptive field size. For example, the \acf{ASPP} in DeepLabV3~\cite{dlv3, dlv3p} uses several parallel dilated~\cite{can} convolution layers to achieve multi-scale feature extraction. Another more recent multi-scale network achieving state-of-the-art 2D image segmentation performance is the HRNet~\cite{hrnet}, which also frequently exchanges multi-scale features extracted through parallel branches. From another perspective, multi-scale networks like DeepLabV3 and HRNet are in essence, ensembles of many sub-architectures allowing multi-scaled features to be aggregated. Through training, the network can decide the optimal combination of features that best suit the data.

 In pursuit of more explicit architecture-level adaptation to arbitrary datasets, works such as nnUNet~\cite{nnunet} introduced a procedural method to self-configure models according to the geometry of the training dataset. The result is a highly-curated model that suits different training datasets, albeit one at a time. In the meantime, there have also been gradient-based self-configuring networks. Fabric-like \acp{CNN}~\cite{grid,interlink,fabric} were proposed to encapsulate many sub-architectures using interlaced multi-scale convolutional blocks, and all the blocks are trained end-to-end using gradient descent. Recently, \ac{NAS} methods like AutoDeepLab~\cite{autodeeplab, c2fnas, rlsearch} have demonstrated more explicit gradient-based self-configuration. It employed weighted trainable connections between cells during training. During training, these weights are tuned to best adapt to the training data. After training, weak connections can be pruned to reveal a compact architecture for the specific training set. The resulting architecture was found to be as successful as many hand-crafted architectures. However, for multi-dataset training, the limitation of these methods is that they are highly curated, and thus the resulting models are not generally applicable to diverse datasets.

\subsubsection{Simultaneous multi-dataset 3D medical image segmentation}
Since most segmentation techniques are transferable between 2D and 3D, 3D medical image segmentation methods have been rapidly evolving by building on works in both 2D and 3D. Some 2D works were first transferred to 3D without major modification~\cite{3dunet, vnet}. Over time, 3D segmentation methods have become increasingly sophisticated ~\cite{unetr, dstunet, tmi1, tmi2}. However, 3D medical image segmentation still needs to combat data scarcity and memory consumption issues. 

Incorporating multiple datasets when training a segmentation model can be a solution to data scarcity. Multi-dataset works such as transfer learning~\cite{transfer1, transfer2, transfer3, transfer4, med3d} and simultaneous multi-dataset learning~\cite{3du2net, heter, cotrain, 3dmdunet} have demonstrated weight-sharing as a practical way to improve performance. However, these methods are limited to a small number of datasets. For example, \cite{heter} uses components developed for a limited scope (CT lesion detection). While \cite{3du2net, 3dmdunet} and \cite{cotrain} demonstrated more flexibility by incorporating domain adapters, they only demonstrated simultaneous segmentation on a small subset of the available datasets (in the \ac{MSD} challenge). Currently, only a small number of works have achieved large-scale applicability to different datasets. 

Memory consumption is bound to affect 3D \acp{CNN} when applied simultaneously to diverse-sized datasets.  In the literature, workarounds such as reducing model complexity~\cite{3dunet} are often required to train 3D \acp{CNN}. However, overly simplifying a \acs{CNN} architecture can limit its multi-dataset learning capacity. Alternatively, patch-based methods~\cite{patch-basic, patch1, patch2, patch3} are often used to divide the large 3D volume into smaller patches. However, this also results in additional hyper-parameters such as patch size, which are dataset-specific. Hence, for simultaneous multi-dataset 3D segmentation, it is important to ensure the model is as end-to-end as possible. 

 \section{Methods}
 \subsection{Network architecture}

\begin{figure*}[htbp]
\centering
    \includegraphics[width=1\textwidth]{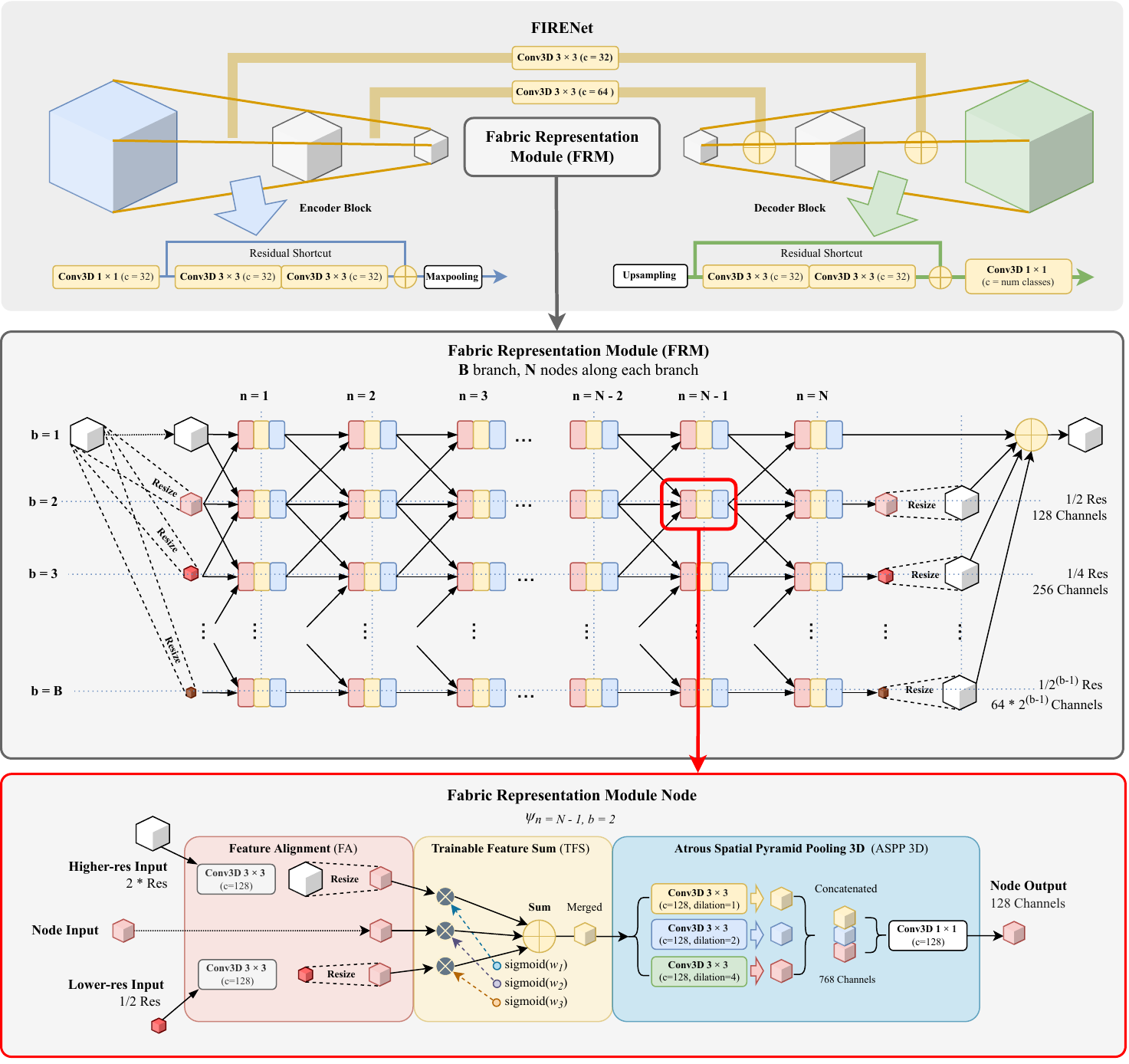}
    \caption{Hierarchical component diagram of \ac{FIRENet} including I) the high-level structure (top), II) the \acf{FRM} (middle), and III) node structure diagram (bottom). \ac{FRM} is designed to be diverse-scale and generalisable for multi-dataset processing. It contains $B$ multi-scale branches and each branch has $N$ nodes. The nodes are densely connected via trainable weights. The nodes perform further multi-scale feature extraction for a fine coverage of different feature sizes.}
    \label{fig:arc:fabric} 
\end{figure*}

\subsubsection{Encoder decoder structure}
\ac{FIRENet} is a 3D \ac{CNN} encoder-decoder with a \acf{FRM}  (Figure~\ref{fig:arc:fabric} top block) in the bottleneck. The encoder performs progressive down-sampling on the input while widening the depth using two residual blocks of 32 and 64 filters, and the decoder mirrors the encoder stricture to recover image resolution. UNet-like shortcuts are used to pass features from the encoder blocks to the corresponding decoder blocks. Each shortcut goes through a $3\times3$ convolutional layer (without altering channel size) to residues semantic gap~\cite{unetpp}. Unlike UNet, element-wise summation is used to merge shortcut features and decoder features.

\subsubsection{Fabric representation module (FRM)}
The latent \acf{FRM} (Figure~\ref{fig:arc:fabric} middle block) is located at the end of the second encoder residual block, and it is parameterised by hyper-parameters $B$ and $N$:
\begin{itemize}
     \item $B$ is the number of branches (from $b=1$ to $b=B$) in \ac{FRM}. The input is tri-linear resized into $B$ scales and processed processed through the $B$ branches, respectively. The first branch $b=1$ maintains all the dimensions of the input. As $b$ increases from 1 to $B$, the spatially dimensions are progressively down-sampled by a factor of ${2}^{b-1}$. In the meantime, the channel size progressively doubles also by a factor or ${2}^{b-1}$ to compensation for the loss of spatial resolution.
     
     \item $N$ is the number of feature extractor nodes along each branch. The nodes are denoted $\psi_{n,\ b}$ (where $n\in\{1...N\}$ and $b\in\{1...B\}$) to indicate their locations in the \ac{FRM}. To build a dense mesh of multi-scale sub-architectures, each node performs multi-scale feature extraction on outputs from its predecessor node (node input) and two neighbouring-branch predecessor nodes. That is, $\psi_{n,\ b}$ takes the outputs from $\psi_{n-1,\ b}$, $\psi_{n-1,\ b-1}$ and $\psi_{n-1,\ b+1}$ as inputs for processing.
\end{itemize}

\subsubsection{Fabric representation module node}
Each node (Figure~\ref{fig:arc:fabric} bottom block) consists of three components to perform multi-scale feature aggregation on the three inputs:
\begin{itemize}
    \item Feature alignment (FA) module to resize the two neighbouring branch inputs (higher and lower resolution) to the size of the node input. First, two 3D convolutions are used to resize the channel dimension. Then, tri-linear up and down-sampling operations are used to resize the spatial dimensions.
    
    \item Trainable feature sum (TFS) module to combine the resized neighbouring inputs with the node input. Inspired by AutoDeepLab's approach to architecture-level adaptability, here, we apply a trainable multiplicative weight to each input. Then, the intensity-scaled inputs are fused using element-wise summation. The three weights are normalised to between [0, 1] using the sigmoid function.
    
    \item \acf{ASPP3D} module to perform multi-scale feature extraction on the fused inputs. Here, three parallel $3\times3$ dilated convolutions with rates 1, 2, and 4 are used. Combined with the multi-branch structure of \ac{FRM}, \ac{ASPP3D} increases the unique receptive field counts by a factor of 3, accounting for a large number of different scaled features. For example, an \ac{FRM} with three branches each using \acs{ASPP3D} nodes with dilation rates 1, 2 and 4 would yield nine unique receptive field sizes (3, 5, 6, 7, 10, 12, 14, 20 and 28). This would otherwise require nine dedicated branches without \acs{ASPP3D}.
\end{itemize}

\begin{figure}[htbp]
\centering
    \includegraphics[width=.48\textwidth]{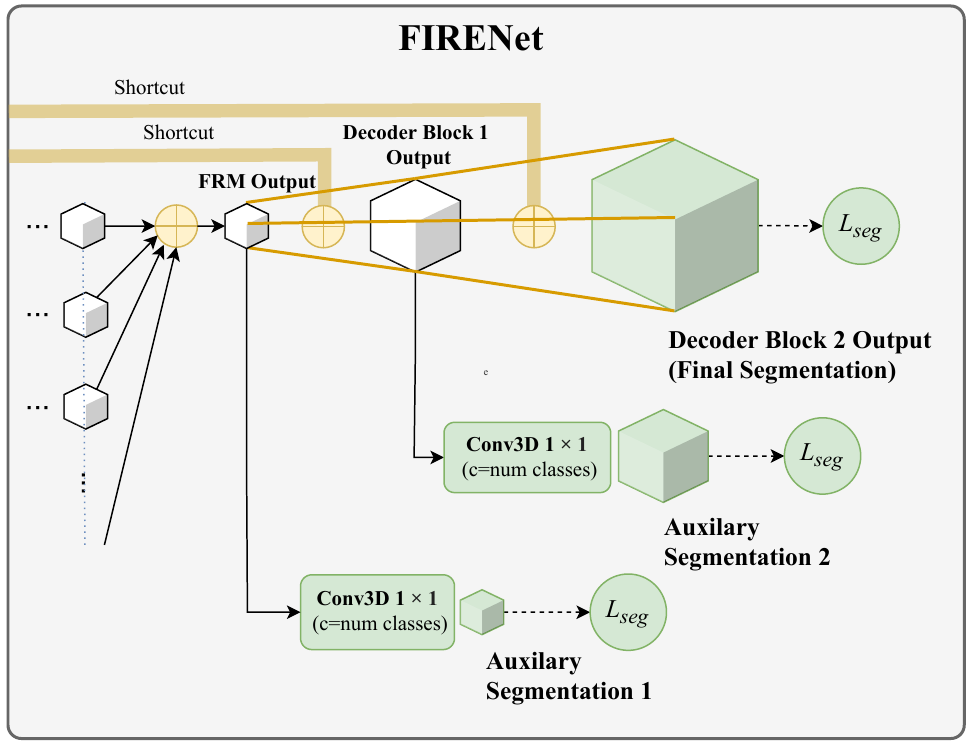}
    \caption{\ac{FIRENet} produces auxiliary segmentation outputs for deep supervision. There are two auxilary outputs and one final output to comptue the final loss.}
    \label{fig:ds} 
\end{figure}

\subsection{Multiple decoders for non-overlapping classes}
Multiple decoders are used when training on datasets with few class overlaps. This is to reduce memory consumption since the segmentation channels cannot be sufficiently shared. As a workaround, each dataset is given a designated decoder, and only one decoder is active at a time. All the decoders are identical (to the decoder in Figure~\ref{fig:arc:fabric}) except for the output channels. Although this workaround is a minor compromise for the otherwise entirely unified \ac{FIRENet}, the majority of the diverse-scaled features are still shared via \ac{FRM}.

\subsection{Training}
The loss function $L_{seg}$ is a combination of the categorical cross-entropy loss and the \ac{DSC} loss. The \ac{DSC} loss was for each channel separately to avoid bias towards majority classes. \ac{FIRENet} used deep supervision~\cite{deepsupervision} during training as it has been shown to improve the performance of encoder-decoder segmentation models~\cite{improvedunet, 3dmdunet}. The outputs from \ac{FRM} and the intermediate decoder blocks are fed through a $1\times1$ 3D convolution to produce auxiliary outputs. A separate segmentation loss $L_{seg}$ is computed for auxiliary output and the total loss is the sum of all the segmentation losses.

The model was trained for 50 epochs using the Adam optimiser~\cite{adam} and a batch size of 1. The hardware used for acceleration was an NVIDIA Tesla V100 (32GB) and the training lasted approximately 36 hours.

\subsection{Experiment setups}
The experiments involve applying \ac{FIRENet} to multiple datasets with diverse image sizes. Since \ac{FIRENet} can process non-uniform image sizes because of its emphasis on diverse-scale feature learning, there was no need to resize all the images to the same size (even for the extreme image sizes).

\subsubsection{Experiment I: Multi-dataset Transfer Learning}
\ac{FIRENet} was tested for transfer learning involving several 3D medical imaging datasets. \ac{FIRENet} was first  pre-trained on a 3D prostate \ac{MR} dataset~\cite{prostate-dataset} (the prostate dataset). Then, the trained \ac{FIRENet} was transferred for simultaneous multi-dataset bone segmentation on a composite bone dataset (the multi-bone dataset). Elastic deformation was used for data augmentation~\cite{unet}. Consistent with the previous CAN3D work~\cite{can3d}, 3-fold validation was used to split the prostate and the multi-bone dataset.

The prostate dataset contains 211 3D \ac{MR} examinations of the pelvic region with manual segmentation labels for five foreground classes: body, bone (pelvic spine and girdle, proximal femur), urinary bladder, rectum and prostate. The multi-bone dataset is composed of four smaller datasets: three 3T \ac{MR} imaging musculoskeletal (MSK) datasets (knee~\cite{mskknee}, shoulder~\cite{mskshoulder}, hip ~\cite{mskhip}) and the OAI knee~\cite{AMBELLAN2019109} dataset. The main difficulty of segmenting the multi-bone dataset is the diverse image sizes and imbalanced numbers of training examples in each dataset (62, 25, 53 and 507 \ac{MR} examinations, respectively). A 3-fold data split was applied to each constinuent dataset. Smaller datasets were duplicated to ensure sample balance.

Due to the lack of existing simultaneous results for the multi-bone segmentation task,  nnUNet was chosen and trained to establish a baseline.  The preprocessing, training and evaluation stages were performed using the official code~\footnote{https://github.com/MIC-DKFZ/nnUNet}. To ensure compatibility with the prescribed training pipeline,  all the substituent bone datasets were pooled into a single dataset for binary bone segmentation.

\begin{table*}[h]
    \centering
    \caption{Pre-training stage segmentation results of \ac{FIRENet} on a MR prostate dataset compared against baseline methods 3D UNet, 3D improved UNet, VNet and CAN3D. The metrics listed are \acf{DSC}, \acf{HD} and mean surface distance (MSD).}
    \label{lbl:prostate-dl}
    \scalebox{0.98}{
    \begin{tabular}{|c|l|ccccc|}
    \hline
    \multicolumn{1}{|c|}{\textbf{Class}} & \textbf{Metrics} & \multicolumn{1}{c}{3D UNet~\cite{unet}} & \multicolumn{1}{c}{3D improved UNet~\cite{improvedunet}} & \multicolumn{1}{c}{VNet~\cite{vnet}} & \multicolumn{1}{c}{CAN3D~\cite{can3d}} & \multicolumn{1}{c|}{\textbf{FIRENet}} \\ \hline
    \multirow{3}{*}{\textbf{Body}}       & DSC (mean $\pm$ sd)                   & 0.96 $\pm$ 0.08           & 0.98 $\pm$ 0.01                    & 0.97 $\pm$ 0.01           & 0.98 $\pm$ 0.01            & \textbf{0.99 $\pm$ 0.01}               \\
                                         & HD (mean $\pm$ sd)               & 42.73 $\pm$ 37.42         & 35.80 $\pm$ 12.02                  & 29.00 $\pm$ 9.34          & 30.57 $\pm$ 11.36          & \textbf{13.08 $\pm$ 6.48}              \\
                                         & MSD (mean $\pm$ sd)               & 3.61 $\pm$ 7.78           & 1.03 $\pm$ 0.69                    & 1.33 $\pm$ 0.55           & 0.95 $\pm$ 0.61            & \textbf{0.47 $\pm$ 0.38}               \\ 
                                         & DSC (min)               & 0.65                      & 0.91                               & 0.92                      & 0.92                       & \textbf{0.93}                                  \\ \hline
    \multirow{3}{*}{\textbf{Bone}}       & DSC (mean $\pm$ sd)                   & 0.91 $\pm$ 0.02           & 0.91 $\pm$ 0.02                    & 0.85 $\pm$ 0.04           & 0.91 $\pm$ 0.02            & \textbf{0.92 $\pm$ 0.02}               \\
                                         & HD (mean $\pm$ sd)               & 36.35 $\pm$ 23.70         & 41.05 $\pm$ 28.78                  & 51.33 $\pm$ 17.70         & 39.73 $\pm$ 24.48          & \textbf{13.19 $\pm$ 11.09}             \\
                                         & MSD (mean $\pm$ sd)               & 1.65 $\pm$ 0.42           & 1.52 $\pm$ 0.39                    & 2.54 $\pm$ 1.15           & 1.50 $\pm$ 0.38            & \textbf{0.85 $\pm$ 0.22}              \\
                                         & DSC (min)               & 0.80                      & 0.85                               & 0.67                      & 0.84                       & \textbf{0.85}                                   \\ \hline
    \multirow{3}{*}{\textbf{Bladder}}    & DSC (mean $\pm$ sd)                   & 0.87 $\pm$ 0.21           & 0.93 $\pm$ 0.10                    & 0.80 $\pm$ 0.16           & 0.92 $\pm$ 0.13            & \textbf{0.93 $\pm$ 0.10}               \\
                                         & HD (mean $\pm$ sd)               & 30.53 $\pm$ 40.06         & 27.94 $\pm$ 35.66                  & 39.61 $\pm$ 36.36         & 33.89 $\pm$ 37.92          & \textbf{9.50 $\pm$ 16.49}              \\
                                         & MSD (mean $\pm$ sd)               & 4.53 $\pm$ 12.69          & 2.10 $\pm$ 3.99                    & 3.91 $\pm$ 3.01           & 2.51 $\pm$ 5.44            & \textbf{1.35 $\pm$ 3.37}               \\
                                         & DSC (min)               & 0.00                      & \textbf{0.27}                               & 0.07                      & 0.06                       & 0.23                                   \\\hline
    \multirow{3}{*}{\textbf{Rectum}}     & DSC (mean $\pm$ sd)                   & 0.78 $\pm$ 0.11           & 0.87 $\pm$ 0.05                    & 0.76 $\pm$ 0.09           & 0.85 $\pm$ 0.05            & \textbf{0.87 $\pm$ 0.05}               \\
                                         & HD (mean $\pm$ sd)               & 27.74 $\pm$ 28.71         & 18.53 $\pm$ 25.34                  & 38.28 $\pm$ 45.40         & 33.30 $\pm$ 36.65          & \textbf{14.14 $\pm$ 20.76}             \\
                                         & MSD (mean $\pm$ sd)               & 3.16 $\pm$ 2.93           & 1.71 $\pm$ 1.10                    & 3.62 $\pm$ 5.95           & 2.08 $\pm$ 1.45            & \textbf{1.24 $\pm$ 1.29}               \\
                                         & DSC (min)               & 0.18                      & \textbf{0.67}                               & 0.29                      & 0.66                       & 0.64                                   \\ \hline
    \multirow{3}{*}{\textbf{Prostate}}   & DSC (mean $\pm$ sd)                   & 0.75 $\pm$ 0.18           & 0.84 $\pm$ 0.08                    & 0.74 $\pm$ 0.13           & 0.81 $\pm$ 0.10            & \textbf{0.86 $\pm$ 0.06}               \\
                                         & HD (mean $\pm$ sd)               & 28.23 $\pm$ 69.67         & 8.91 $\pm$ 9.17                    & 25.38 $\pm$ 29.83         & 14.28 $\pm$ 16.33          & \textbf{7.36 $\pm$ 10.10}              \\
                                         & MSD (mean $\pm$ sd)               & 15.85 $\pm$ 65.41         & 1.92 $\pm$ 0.87                    & 3.54 $\pm$ 2.89           & 2.17 $\pm$ 0.99            & \textbf{1.13 $\pm$ 0.48}               \\ 
                                         & DSC (min)               & 0.00                      & 0.45                               & 0.08                      & 0.39                       & \textbf{0.51}                                  \\\hline
    \end{tabular}}
\end{table*}

\begin{table*}[t]
    \centering
    \caption{Pre-training stage segmentation results of \ac{FIRENet} on a MR prostate dataset compared with published traditional baselines. \ac{FIRENet} used 3-fold validation and the traditional methods used leave-one-out. The metrics listed are \acf{DSC}, \acf{HD} and mean surface distance (MSD).}
    \label{lbl:prostate}
    \scalebox{0.80}{
    \begin{tabular}{|c|cccccccc|}
    \hline
    \textbf{Method}                        & \begin{tabular}[c]{@{}c@{}}Median\\ Body DSC\end{tabular} & \begin{tabular}[c]{@{}c@{}}Median\\ Bone DSC\end{tabular} & \begin{tabular}[c]{@{}c@{}}Median\\ Bladder DSC\end{tabular} & \begin{tabular}[c]{@{}c@{}}Median\\ Rectum DSC\end{tabular} & \begin{tabular}[c]{@{}c@{}}Median\\ Prostate DSC\end{tabular} & \begin{tabular}[c]{@{}c@{}}Median\\ Prostate MSD (mm)\end{tabular} & \begin{tabular}[c]{@{}c@{}}Median \\ Prostate HD (mm)\end{tabular} & \begin{tabular}[c]{@{}c@{}}Mean\\ Prostate DSC\end{tabular} \\ \hline
    Dowling et al.\cite{DOWLING20151144}                & \textbf{1}                                                & 0.92                                                      & 0.86                                                         & 0.85                                                        & 0.82                                                          & 2.04                                                          & 13.3                                                          & 0.80                                                           \\
    Chandra et al.\cite{Chandra_2016}                 & 0.94                                                      & 0.81                                                      & 0.87                                                         & 0.79                                                        & 0.81                                                          & 2.08                                                          & 9.60                                                          & 0.79                                                           \\
   \textbf{FIRENet} & 0.99                                                      & \textbf{0.92}                                             & \textbf{0.96}                                                & \textbf{0.89}                                               & \textbf{0.87}                                                 & \textbf{1.00}                                                 & \textbf{4.24}                                                 & \textbf{0.86}                                                  \\ \hline
    \end{tabular}}
\end{table*}

\subsubsection{Experiment II: Simultaneous multi-dataset segmentation on \ac{MSD}}
\ac{FIRENet} was used to simultaneously segment the 10 datasets from the \ac{MSD} model generalisability challenge. Most existing simultaneous multi-dataset segmentation works (\cite{3du2net} and \cite{cotrain}) were performed on a subset of these 10 datasets, excluding important and challenging tasks like HepaticVessel and Lung. The current work evaluated \ac{FIRENet} on all 10 \ac{MSD} datasets to provide a more thorough evaluation. These 10 datasets are highly diverse in image size and voxel spacing, with the smallest dimension being 11 voxels and the largest dimension being 751 voxels. For preprocessing, each image was re-sampled to the same voxel spacing of [1, 1, 1] (but not the same size) to retain the relative object sizes. Cropping was used sparing only to cap large image dimensions to 180 voxels (which covers most of the image after re-sampling). Finally, voxel intensity standardisation was applied before entering the network. The dataset was divided into training and validation sets according to an 80\%-20\% split ratio as per previous work~\cite{3dmdunet}.

Since the 10 datasets have almost no class overlap, 10 decoders are used to produce segmentation maps for the 10 datasets, respectively. The training process iterates through the 10 datasets activating one decoder at a time (according to the dataset). The \ac{FRM}, which contains learnt multi-scale features, was shared the whole time.

\begin{figure*}[t]
    \centering
    \includegraphics[width=.95\textwidth]{images/convergence.pdf}
    \caption{Convergence plots comparing \acs{FIRENet}-R (randomly initialised) and FIRENet-T (pre-trained on the prostate dataset and fine-tuned on the four MSK bone datasets). The \acp{DSC} for the bones were averaged across the multi-bone dataset. FIRENet-T consistently showed accelerated convergence, especially in the first 30 epochs.}
    \label{fig:convergence}
\end{figure*}

\section{Results and discussion}

\subsection{Experiment I: Multi-dataset Transfer Learning}
\subsubsection{\ac{FIRENet} pre-training evaluation}
Table~\ref{lbl:prostate-dl} shows the subject-level, 3-fold validation segmentation results from the prostate MR pre-training experiment for the body, bone, urinary bladder, rectum and prostate classes. \ac{FIRENet} was compared to four other 3D deep learning baseline methods (3D UNet~\cite{3dunet}, improved UNet~\cite{improvedunet}, VNet~\cite{vnet} and CAN3D~\cite{can3d}) on the same data splits. Overall \ac{FIRENet} produced better results than these baseline methods in \ac{DSC}, \ac{HD} and mean surface distance values across multiple classes. In terms of the performance on outlier cases (min \ac{DSC}), \ac{FIRENet} was amongst the most resilient models producing less severe segmentation errors (indicated by the higher min \acp{DSC} in the table).

\begin{figure*}[t]
    \centering
    \includegraphics[width=.95\textwidth]{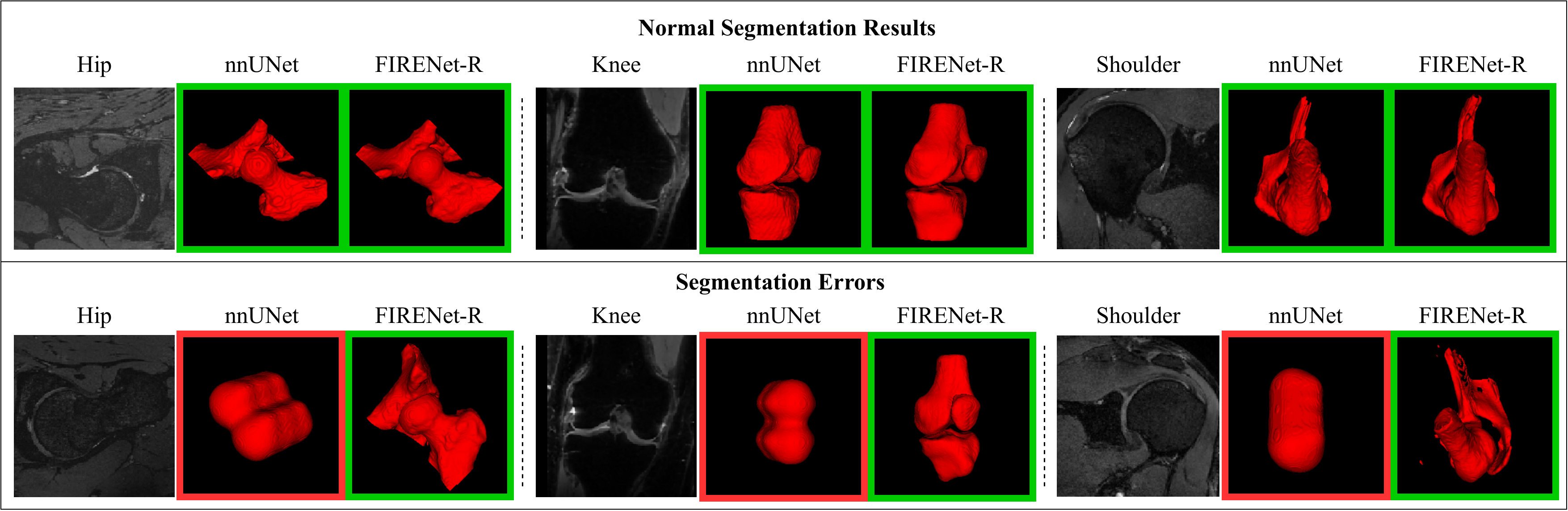}
    \caption{Comparison of \ac{FIRENet} and nnUNet when handling multiple datasets with diverse image sizes. \ac{FIRENet} appeared more resilient, producing valid results in all three cases (and also for the entire dataset) and nnUNet appeared to have struggled to be consistent. Both methods used the same data splits, and nnUNet was built and trained using the official nnUNet scripts.}
    \label{fig:nnunetfail}
\end{figure*}

\begin{figure}[h!]
     \centering
     \includegraphics[width=0.45\textwidth]{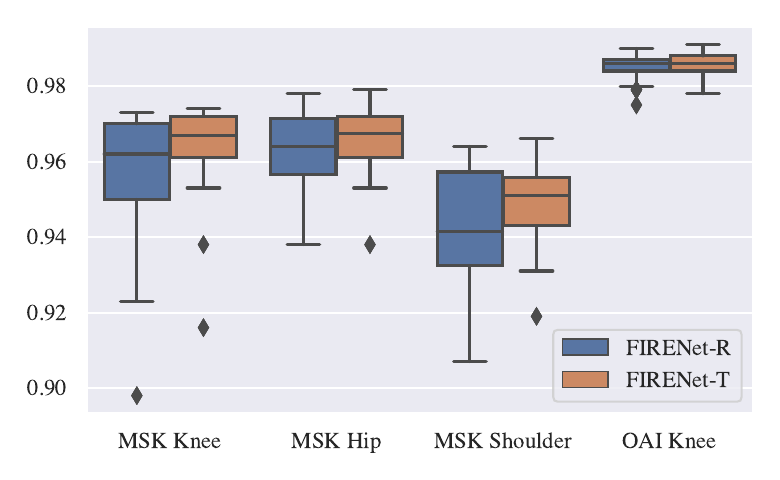}
     \caption{Box plot comparing \ac{FIRENet}-R  and \ac{FIRENet}-T results on the multi-bone dataset. The corresponding quantitative results can be found in Table~\ref{lbl:bone}.}
     \label{fig:box}
\end{figure}

\begin{table}[h]
    \centering
    \caption{Mean bone segmentation \acfp{DSC} of FIRENet-R (randomly initialised) and FIRENet-T (pre-trained on a prostate dataset and
fine-tuned on the four MSK bone datasets) compared to other baseline methods. nnUNet was trained by treating all four bone datasets as one. \ac{FIRENet} appeared more adaptable when applied to multiple datasets with non-uniform image sizes. Whereas nnUNet was prone to segmentation failure (see examples in Figure~\ref{fig:nnunetfail}). The cited previous methods only applied to and reported single-dataset results.}
    \label{lbl:bone}
\scalebox{0.95}{
\begin{tabular}{|l|cccc|c|}
\hline
Method  & \multicolumn{1}{l}{Hip} & \multicolumn{1}{l}{Knee} & \multicolumn{1}{l}{Shoulder} & \multicolumn{1}{l|}{OAI} & \multicolumn{1}{l|}{Overall} \\ \hline
FIRENet-R  & 0.954 & 0.961 & 0.940 & 0.985 & 0.952 \\
FIRENet-T  & \textbf{0.962} & \textbf{0.965} & \textbf{0.947} & \textbf{0.986} & \textbf{0.965} \\ 
\begin{tabular}[c]{@{}l@{}}nnUNet \end{tabular} & 0.612 & 0.851 & 0.633 & 0.986 & - \\ \hline
\begin{tabular}[c]{@{}l@{}}CAN3D  \cite{can3d}\end{tabular}  & - & - & - & 0.986 & - \\
\begin{tabular}[c]{@{}l@{}}SSM+CNN  \cite{AMBELLAN2019109}\end{tabular}   & - & -  & - & 0.985 & - \\ \hline
\begin{tabular}[c]{@{}l@{}}SSM  \cite{mskhip}\end{tabular} & 0.950  & - & - & - & - \\
\begin{tabular}[c]{@{}l@{}}SSM \cite{mskknee}\end{tabular} & - & \begin{tabular}[c]{@{}c@{}}0.922\\

\end{tabular} & - & - & - \\
\begin{tabular}[c]{@{}l@{}}SSM  \cite{mskshoulder}\end{tabular} & - & - & \begin{tabular}[c]{@{}c@{}}0.8815\\\end{tabular} & - & - \\ \hline
\end{tabular}}
\end{table}

Table \ref{lbl:prostate} shows results from published baseline methods~\cite{DOWLING20151144, Chandra_2016} on this prostate dataset. These results are presented separately as they are from statistical methods. Nonetheless, their quantitative results are comparable even to the more recent deep learning results (UNet and VNet), likely because deep learning requires a lot of training data. \ac{FIRENet} and CAN3D were the only two methods significantly exceeding these non-deep-learning methods, especially across the most challenging classes (prostate and rectum). 

\subsubsection{Transfer learning}
For the transfer learning part of the experiment, two \ac{FIRENet} instances were trained on the multi-bone dataset: a \ac{FIRENet} with weights pre-trained on the prostate dataset  (FIRENet-T) and a randomly initialised \ac{FIRENet} (FIRENet-R). The multi-bone segmentation results of FIRENet-R and FIRENet-T are compared in Tables \ref{lbl:bone} and Figure~\ref{fig:box}. Overall, both \ac{FIRENet}-R and T exhibited the ability to simultaneously segment diverse medical image data despite relying on only one shared set of weights: their results on the OAI dataset are comparable to other methods~\cite{can3d, AMBELLAN2019109} which published results optimised for only one dataset. Comparing the mean \ac{DSC} values of \ac{FIRENet}-R and \ac{FIRENet}-T, it can be seen that \ac{FIRENet}-T could leverage pre-training to improve the segmentation performance across all the constituent bone datasets. Moreover, as Figure~\ref{fig:box} shows, \ac{FIRENet}-T's lowest \ac{DSC} results were also noticeably improved over \ac{FIRENet}-R, indicating reduced critical segmentation errors. In terms of convergence, \ac{FIRENet}-T was also more stable and faster (Figure~\ref{fig:convergence}) supporting the benefits of transfer learning for simultaneous multi-dataset segmentation.

\begin{figure*}[h]
     \centering
     \includegraphics[width=\textwidth]{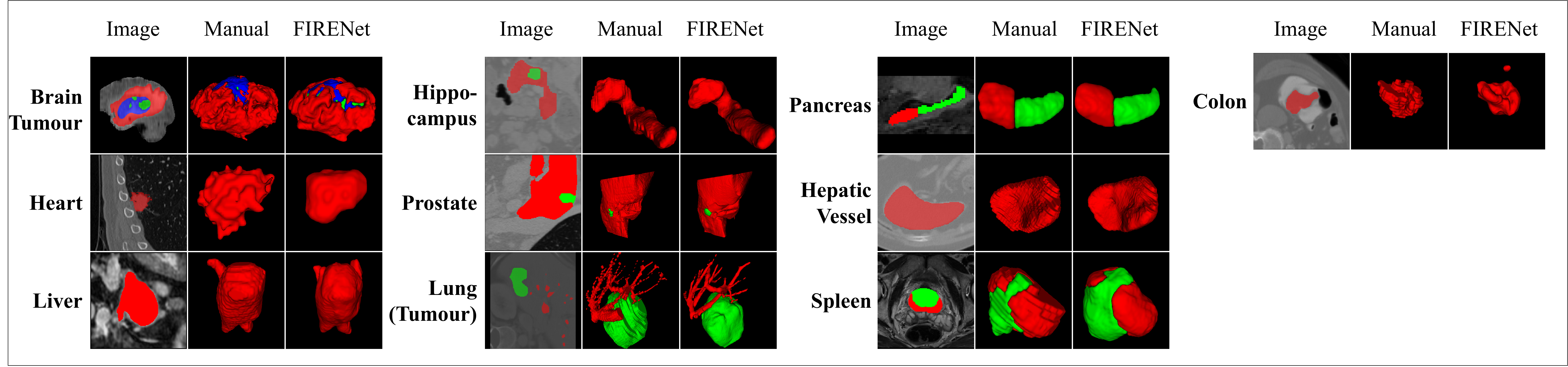}
     \caption{\ac{FIRENet}'s simultaneous segmentation results on the 10 \ac{MSD} datasets. The image dimensions are highly diverse in size with a difference of up to 4 times for each dimension.}
     \label{fig:msdall}
\end{figure*}

\begin{table*}[h]
\caption{FIRENet's simultaneous segmentation \ac{DSC} results on the full \ac{MSD} dataset. The baseline methods for comparison are 3D MDUNet~\cite{3dmdunet} and 3D U$^2$Net. All of the \ac{DSC} results were computed without the background class.}
\centering
\label{tbl:III}
\begin{tabular}{|l|c|c|c|c|}
\hline
Task Code             & \ac{FIRENet}                           & 3D MDUNet  & 3D MDUNet(3D U$^2$Net baseline)  & 3D U$^2$Net(universal) \\ \hline
Task01\_BrainTumour   & 0.571                                   & -         & -                                 & -\\
Task02\_Heart         & 0.913                                   & 0.921     & 0.913                             & 0.919\\
Task03\_Liver         & 0.917                                   & -         & -                                 & 0.935\\
Task04\_Hippocampus   & 0.826                                   & 0.650     & 0.624                             & 0.882\\
Task05\_Prostate      & 0.807                                   & -         & -                                 & 0.789\\
Task06\_Lung          & 0.778                                   & -         & -                                 & -\\
Task07\_Pancreas      & 0.742                                   & 0.622     & 0.534                             & 0.621\\
Task08\_HepaticVessel & 0.787                                   & -         & -                                 & -\\
Task09\_Spleen        & 0.895                                   & 0.833     & 0.815                             & -\\
Task10\_Colon         & 0.437                                   & -         & -                                 & -\\ \hline
\end{tabular}
\end{table*}

nnUNet was chosen and trained to establish a baseline for the multi-bone segmentation experiment. However, unlike \ac{FIRENet}, the multi-bone segmentation results from nnUNet indicate that it is less suitable for simultaneous multi-dataset processing - the best nnUNet \acp{DSC} were substantially lower at 0.612, 0.851, 0.633 and 0.986 for the MSK hip, knee, shoulder and OAI, respectively. In this case, nnUNet produce inconsistent segmentation result. As the segmentation visualisation in Figure~\ref{fig:nnunetfail} shows, there were unpredictable failures in nnUNet's results on three of the four constituent datasets (MSK knee, shoulder and hip). It was also observed that the training process of nnUNet was highly unstable, and early stopping was required to obtain some usable results (which were already presented in this section). nnUNet's inability to produce consistent and satisfactory results could be attributed to its one-size-fits-all self-configuring procedure, as well as the lack of rich-scale feature extraction as desired by multi-dataset applications. On the contrary, FIRENet-R and FIRENet-T were designed to be voided of dataset-specific configurations, there is also a strong emphasis on multi-dataset feature learning to cover different data sizes.

\subsection{Experiment II: Simultaneous multi-dataset segmentation on \acf{MSD}}
In this experiment, FIRENet's segmentation performance was compared to the recently published simultaneous multi-dataset segmentation methods (3D MDUNet~\cite{3dmdunet} and 3D U$^2$Net), and some example predictions are visualised in Figure~\ref{fig:msdall}. As \ref{tbl:III} shows, compared to 3D MDUNet and its 3D U$^2$Net baseline (as reported in the 3D MDUNet publication), \ac{FIRENet}'s \ac{DSC} results in the Hippocampus, Pancreas and Spleen segmentation tasks are considerably improved. Compared to the original 3D U$^2$Net, \ac{FIRENet} achieved significantly better Pancreas segmentation \ac{DSC} (by 12.1\%), and similar segmentation \acp{DSC} in the Heart, Liver and Prostate segmentation tasks. However, it did also under-perform in terms of Hippocampus segmentation \ac{DSC}. It is worth noting that both the 3D MDUNet and 3D U$^2$Net only involved a subset of the \ac{MSD} challenge and, by extension, only a subset of the challenge's complexity. In the meantime, \ac{FIRENet} was trained on all 10 datasets.

While nnUNet has been the gold standard in the \ac{MSD} challenge, as experiment I shows, pooling different datasets together is not within the scope of nnUNet's design. Although nnUNet's results on individual ac{MSD} datasets have been state-of-the-art, they were all specially tuned for one dataset at a time. Hence, nnUNet's single dataset results were not included in the performance comparison with other methods for this experiment. 

\subsection{Architecture-level generalisability}

\ac{FIRENet} exhibited excellent architecture-level generalisability as it can seamlessly transfer between and apply to multiple datasets. The two experiments reported cover applications in various scenarios including single dataset learning, transfer learning and simultaneous multi-dataset learning. Despite the significant differences in image sizes, object appearances and number of samples, No hyper-parameter modification was required between experiments showing the benefits of an all-inclusive fabric architecture. The only workaround required was the use of multiple decoders to save memory (in the \ac{MSD} segmentation experiment). 

\subsection{Limitations}
As \ac{FIRENet} universally applies to different imaging datasets, its design and training methodology purposefully lack domain-specific optimisations. Compared to public challenge results produced by highly specialised and single-dataset methods, a general architecture does not often yield optimal numerical results across all the different datasets. Although \ac{FIRENet}'s architecture is free of overhead in handling multiple datasets, there is an increase in total model size with each additional decoder (required for different output formats). Finally, as a \ac{CNN}, \ac{FIRENet} faces the current limitations of deep learning. For example, the lack of clinical explain-ability and difficulty extrapolating to unseen data distributions.

\subsection{Future work}
As new labelled medical imaging datasets and deep learning accelerators become available, \ac{FIRENet}'s size and training set composition can continue to expand. It also would be beneficial to train multiple instances of \ac{FIRENet} to specialise in different imaging modalities or anatomical structures. In terms of architectural development, \ac{FIRENet} could be effectively re-purposed for classification, regression and multi-task learning by adding downstream prediction heads. For example, a classification head could be attached to the output of \ac{FRM} for 3D medical image classification based on the features extracted.

\section{Conclusion}
To better take advantage of the scattered dataset in medical image analysis, this work presents \ac{FIRENet}, a versatile 3D neural network architecture geared towards simultaneous multi-dataset learning and segmentation. In the prostate, multi-bone and \ac{MSD} segmentation experiments, \ac{FIRENet} exhibited superior multi-dataset adaptability without dataset specific tuning or even hyper-parameter modification. In contrast, existing works in the literature either performed poorly across multiple datasets or were limited to a small number of datasets. The technical contribution of \ac{FIRENet} is the multi-scale \acf{FRM}, which acts as a super-position of many possible sub-architectures alleviating the need for dataset-specific architecture design. In addition, node in \ac{FRM} employs \ac{ASPP} 3D for further diverse-scale feature extraction, which ensures fine-grained coverage of different scaled features. For future work, \ac{FIRENet}'s \ac{FRM} has the potential to serve as a backbone feature extractor for other 3D medical image analysis tasks.

\bibliographystyle{plain}
\bibliography{ref}

\acrodef{CNN}{{Convolutional Neural Network}}
\acrodef{FIRENet}{{Fabric Image Representation Encoding Network}}
\acrodef{ASPP}{{atrous spatial pyramid pooling}}
\acrodef{MR}{{magnetic resonance}}
\acrodef{CT}{{computed tomography}}
\acrodef{FRM}{{fabric representation module}}
\acrodef{TFS}{trainable feature sum}

\acrodef{DSC}{{Dice similarity coefficient}}
\acrodef{GPU}{{Graphics Processing Unit}}
\acrodef{DNN}{{Deep Neural Network}}
\acrodef{SSM}{{Statistical Shape Modelling}}
\acrodef{ASPP3D}{{Atrous Spatial Pyramid Pooling 3D}}
\acrodef{MIA}{{medical image analysis}}
\acrodef{LiTS}{{Liver Tumor Segmentation Challenge}}
\acrodef{BraTS}{{Brain Tumor Segmentation Challenge}}
\acrodef{HD}{{Hausdorff distance}}
\acrodef{NAS}{{Neural Architecture Search}}
\acrodef{MSD}{{Medical Segmentation Decathlon}}

\end{document}